\documentclass[showpacs,aps,pre,twocolumn,epsfig]{revtex4}
\usepackage{graphicx}
\usepackage{amsfonts}
\usepackage{amssymb}

\begin{document}
\title{Quasi 1D Bose-Einstein condensate flow past a nonlinear barrier}
\author{F. Kh. \  Abdullaev\dag\ddag \footnote[1]{Corresponding author fatkhulla@yahoo.com} R. M.\ Galimzyanov\ddag
\ and \ Kh.N.\ Ismatullaev\ddag} \affiliation{\dag \ CFTC,
Complexo Interdisciplinar, Universidade Lisboa, Portugal\\
\ddag \ Physical-Technical Institute of the Academy of Sciences,
Bodomzor Yoli street 2-b, 100084, Tashkent-84, Uzbekistan}

\begin{abstract}
 The problem of a quasi 1D {\it repulsive} BEC flow past
through a nonlinear barrier is investigated. Two types of
nonlinear barriers are considered, wide and short range ones.
Steady state solutions for the BEC moving through a wide repulsive
barrier and critical velocities have been found using
hydrodynamical approach to the 1D Gross-Pitaevskii equation. 
It is shown that in contrast to the linear barrier case, for a
wide {\it nonlinear} barrier an interval of velocities $0 < v <
v_-$ {\it always} exists, where the flow is superfluid  regardless
of the barrier potential strength. For the case of the $\delta$
function-like barrier, below a critical velocity two steady
solutions exist, stable and unstable one. An unstable solution is
shown to decay into a gray soliton moving upstream and a stable
solution. The decay is accompanied by a dispersive shock wave propagating downstream
in front of the barrier.
\end{abstract}
\maketitle

\newpage

\section{Introduction}
The problem of the transcritical flow of a BEC through the
penetrable barriers has been under recent active investigations
~\cite{Kamchatnov,Hakim,AP04,EA07}. The damping processes for the
superfluid flow moving through the barrier are of a fundamental
interest. In multidimensional case above some critical velocity of
the obstacle motion the damping accompanied by the radiation
emission ~\cite{AP04} is observed. Thus in the region
when the motion is still superfluid, the velocity is bounded
above. The damping is associated with the Landau type damping and
related to the emission of the elementary excitations. Landau
damping can be described in the framework of the mean field theory
and is not associated with thermalization processes
~\cite{Pitaevskii}. The critical velocity value at which the
damping is observed, differs essentially from the values predicted
by the Landau theory. As it was shown firstly by Feynman
~\cite{Feynman}, the reason is in the nonlinearity of the system.
In the case of a quasi 1D Bose-Einstein condensate flow, when
passing through a penetrable barrier, some interval of velocities
$v_{-}<v<v_{+}$ exists, where  trains of dark solitons are
generated, that leads to deviation from predictions based on the
matching with the spectrum of elementary linear
excitations~\cite{Law,Kamchatnov}.In addition in this range
of velocities, generation of dispersive shock waves occurs.
Experimental proof of the existence of the velocities interval was
given in the work ~\cite {EA07}. Hakim ~\cite{Hakim2} has
indicated that for supersonic velocities (including ones above
supercritical velocity $v_+$) some radiation is still nonzero and
its amplitude   rapidly decreases at the ratio of the
potential variation length to the GPE coherence length. The
amplitude of the wake can be  characterized by the Fourier
transform of the obstacle potential~\cite{Pavloff}. Thus,
wide and smooth potentials can be considered as radiationless at
velocities above {\it supercritical}. Seemingly in one dimensional
case only stable dark solitons can exist. Peculiarity of the one
dimension is in the fact that generation of the solitons is
possible till some {\it supercritical} velocity, $v_+$. Above this
velocity the emission is strongly damped and the
quasi-superfuidity is restored. The radiation exists, but
exponentially small-decay rate is proportional to $l_{pot}/l_h$,
where $l_{h}$ is the healing length of the order of the dark
soliton width.

In this work we consider the phenomena occurring in the flow of a
quasi 1D BEC past {\it a nonlinear}
barrier which is a localized space inhomogeneity of the the
nonlinearity coefficient in the Gross-Pitaevskii equation. Such a
type of barriers can be formed by some area of BEC where the
effective value of the atomic scattering length is varied in {\it
the space}. It can be achieved both by the Feshbach resonance
techniques~\cite{Inouye}, and by the local variation of the
transverse frequency of the trap potential. In the former case,
varying external magnetic field in space near the resonance, one
can vary the value of the atomic scattering length $a_s$.
Another way is to use optically induced Feshbach
resonances~\cite{Schlyapnikov}. In this case the variation can be
achieved by local change in the intensity of a laser field.
Variation of $a_s$ in a half space recently has been suggested to
generate vortices in BEC as a nonlinear piston method~\cite{BG10}.

The present paper is motivated by the works
~\cite{Hakim,Kamchatnov} where flow of a BEC past an obstacle in
one dimension was investigated. We consider two cases, wide
obstacle potential and short range one.

\section{The model}
Let us consider a nonlinear penetrable barrier moving through the
elongated BEC.  A quasi one dimensional BEC can be described by
the Gross-Pitaevsky (GP) equation with standard dimensionless
variables
\begin{equation}
i\psi_t + \frac{1}{2}\psi_{xx} - |\psi|^2\psi =
V(x+vt)|\psi|^2\psi, \label{eq1}
\end{equation}
where
\begin{eqnarray}
t=T\omega_{\perp}, \ x=X/l_{\perp}, \
\psi(x,t)=\sqrt{2|a_s|}\Psi(x,t), \nonumber  \\ l_{\perp}=
\sqrt{\hbar/m\omega_{\perp}} ,  \label{eq2}
\end{eqnarray}
$a_s$ is the atomic scattering length, $\omega_{\perp}$ is the
transverse frequency of the trap, $V\rightarrow \frac{a_s}{a_{s0}}
$, $a_{s0}$ is the background value of the scattering length
$a_s$. For the further study of the flow problem it is useful to
pass to the
reference frame moving with the barrier $ x^{\prime} = x + vt, \ t
= t$. So we come to the equation
\begin{equation}
i\psi_t + iv\psi_{x^{\prime}} + \frac{1}{2}\psi_{x^{\prime}x^\prime} - |\psi|^2\psi =
 V(x^{\prime})|\psi|^2\psi.
 \label{eq3}
\end{equation}
 The scattering length can be  manipulated  with a laser field
tuned near a photo association transition, e.g., close to the
resonance of one of the bound $p$ levels of the excited molecules.
Virtual radiative transitions of a pair of interacting atoms to
this level can change the value and even  reverse the sign of the
scattering length~\cite{Schlyapnikov}. Recently spatial
modulations of the atomic scattering length by the optical
Feshbach resonance method  was realized experimentally in
BEC~\cite{NOLExp2}. Such approach implies some spontaneous
emission loss which is inherent in the optical Feshbach resonance
technique. Here we assume that such dissipative  effects can be
ignored, since they become possible if one uses  laser fields of
sufficiently high intensity detuned from the resonance. Thus the
repulsive nonlinear barrier can be formed by an focused external
laser beam with the parameters lying near the optically induced
Feshbach resonance.

\subsection{ Wide obstacle potential }
We analyze this case following the method developed in
~\cite{Hakim,Kamchatnov} for the linear barrier case. Let us pass
to the hydrodynamical form for the GP equation
(\ref{eq1}). It can be obtained by the following transformation
\begin{equation}
\psi(x^{\prime},t)= \sqrt{\rho(x^{\prime},t)}e^{i\int^{x^\prime}
u(x,t)dx}. \label{eq4}
\end{equation}
Substituting it into (\ref{eq1}) and introducing $u'=u+v$ we
obtain the system
\begin{eqnarray}
\rho_t + (\rho u^{\prime})_{x^{\prime}}=0,\\
u^{\prime}_t + u^{\prime}u^{\prime}_{x^{\prime}} + \left(
\frac{\rho_{x^{\prime}}^{2}}{8\rho^2}
- \frac{\rho_{x^{\prime}x^{\prime}}}{4\rho}\right)_{x^{\prime}}+\nonumber\\
\rho_{x^{\prime}} + (V(x^{\prime})\rho)_{x^{\prime}} = 0.
\label{eq5}
\end{eqnarray}
For a wide smooth obstacle potential we can neglect
the terms in the bracket in the second equation that corresponds
to the hydrodynamical approximation. Omitting also primes, for
stationary solutions we can put $\rho_t=0$ and $u_t=0$, and obtain
the following system of equations
\begin{eqnarray}
(\rho u)_x = 0, \label{difpu1}
\end{eqnarray}
\begin{eqnarray}
 uu_x + \rho_x + (V\rho)_x = 0, \label{difpu2}
\end{eqnarray}
with the boundary conditions
\begin{eqnarray}
\rho \rightarrow 1, \ u \rightarrow v, \ V(x) \rightarrow 0, \
\mbox{when} \ |x| \rightarrow \infty.  \label{bound}
\end{eqnarray}
Integrating over $x$ we find
\begin{equation}
\rho u = v,\label{pu1}
\end{equation}
\begin{equation}
 \frac{1}{2}u^2 + \rho + V(x)\rho = \frac{1}{2}v^2 +
1. \label{pu2}
\end{equation}
Eliminating the function $\rho$ from these equations, we get
\begin{equation}
V(x)=\frac{1}{2v}(u-v)[2-u(u+v)]\equiv F(u). \label{Fu}
\end{equation}
Since we consider repulsive obstacle potential $V(x)>0$ we have
the condition $F(u)>0$. Maximum of $F(u)$ is realized at $u_m =
\sqrt{(v^2 + 2)/3} $. Thus the maximum of the function $F(u)$ is
\begin{equation}
\max[F(u)] = \mu(v) =
\frac{1}{v}\sqrt{\left(\frac{v^2+2}{3}\right)^{3}} - 1.
\label{eqMaxFu}
\end{equation}

Stationary solution $u(x)$ is obtained by solving the equation
(\ref{Fu}) with respect to $u$. This equation has a real solution
defined for all $x$ provided that
\begin{equation}
V_{m }\equiv \max[V(x)] \leq  \max[F(u)], \label{MaxVMaxFu}
\end{equation}
i.e. the range of values of $V(x)$, which is $[0,V_m]$, lies
within the range of values of the function $F(u)$
~\cite{Kamchatnov}.
\begin{figure}[h]
\centerline{\includegraphics[width=8cm]{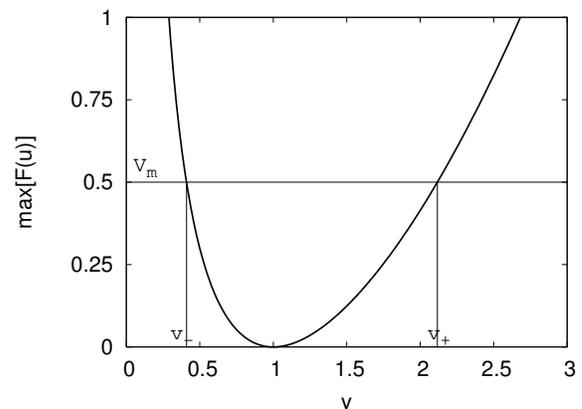}}
\caption{Maximum of the function $F(u)$ (see Eq.(\ref{eqMaxFu}))
versus $x$. For given obstacle potential maximum $V_{m}=0.5$,
critical values of the velocity $v_{-}=0.409$,\ $v_{+}=2.117$}.
\label{maxFu}
\end{figure}

Maximum of the function $F(u)$ versus the obstacle velocity $v$ of
BEC is presented in Fig.~\ref{maxFu}. As seen for any value of
$V_{m}$ two critical values of the velocity exist, $v_{-}, v_+$,
determined by equation $V_{m }=\mu(v)$. In transcritical regime,
in the interval $v_{-} < v < v_+ $, the condition of the
stationary flow (\ref{MaxVMaxFu}) does not hold. Out of this
region, in subcritical ($v<v_{-}$) and supercritical ($v>v_{+}$)
regimes the radiation phenomena are negligible and the motion of
the system can be considered as superfluid.

Analyzing expression (\ref{eqMaxFu}) and Fig.~\ref{maxFu} it
should be noted that unlike the case of a wide linear barrier,
considered in~\cite{Kamchatnov}, the velocity $v_{-}$ is not
vanish and there always exists an interval $0 < v < v_-$ where the
flow is superfluid.

Eq.(\ref{Fu}) can be rewritten as
\begin{equation}
u^3-(v^2+2)u+2v(V(x)+1)=0, \label{u3}
\end{equation}
which is a cubic equation with respect to $u(x)$. Solving it we
obtain the following solutions for $u(x)$ satisfying the boundary
conditions
\begin{eqnarray}
u(x)=-2\sqrt{q}\cos{\left(s(x)-\frac{2\pi}{3}\right)} \qquad
\mbox{for} \  v < v_-, \ \label{sol_u-} \\
u(x)=-2\sqrt{q}\cos{\left(s(x)+\frac{2\pi}{3}\right)} \qquad
\mbox{for} \  v_+ < v,  \label{sol_u+}
\end{eqnarray}
where
$$
q=\frac{v^2+2}{3},\
s(x)=\frac{1}{3}\arccos(\frac{v(V(x)+1)}{\sqrt{q^3}}).
$$
\begin{figure}[h]
\centerline{\includegraphics[width=8cm]{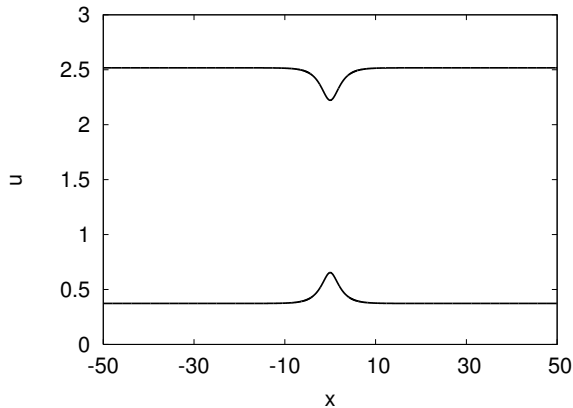}}
\caption{Spatial profiles of the local velocity $u(x)$. The
barrier velocities are equal to $v = 0.373$ and $v = 2.517$ for
lower and upper lines respectively.} \label{ux}
\end{figure}
Spatial profiles of the local velocity $u$ for subcritical
$v=0.373$ ($v<v_{-}$) and supercritical $v=2.517$ ($v>v_{+}$)
regimes are depicted in Fig.~\ref{ux}. The NL obstacle potential
is taken in the form $V(x)=V_{m}/\cosh(x/2)$ with its maximum
value $V_{m}=0.5$.

\begin{figure}[h]
\centerline{\includegraphics[width=8cm]{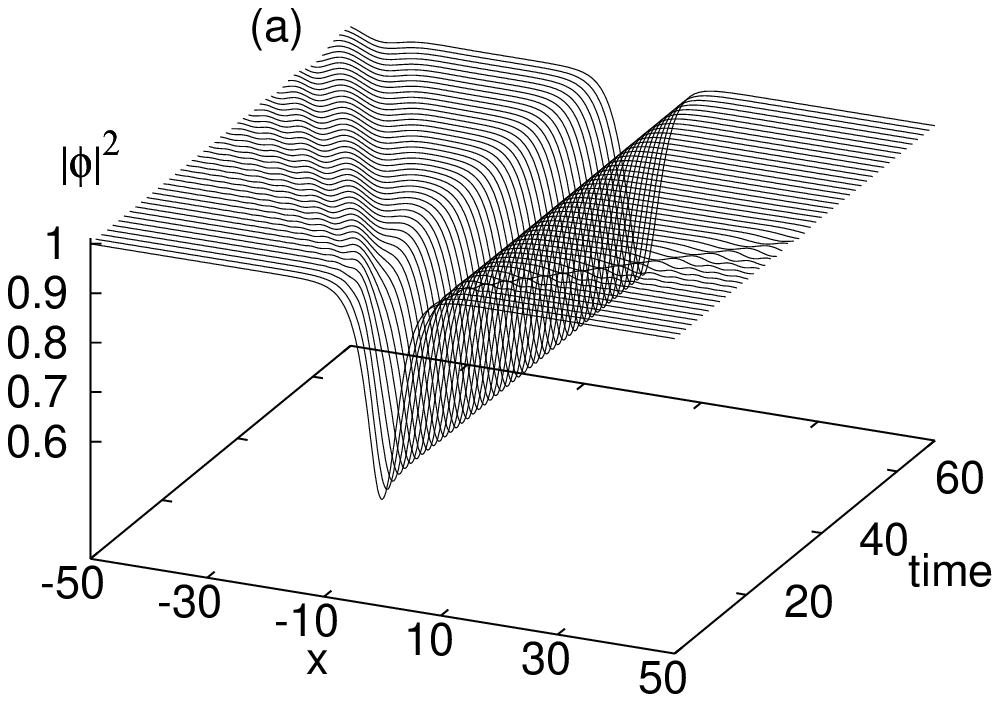}}
\centerline{\includegraphics[width=8cm]{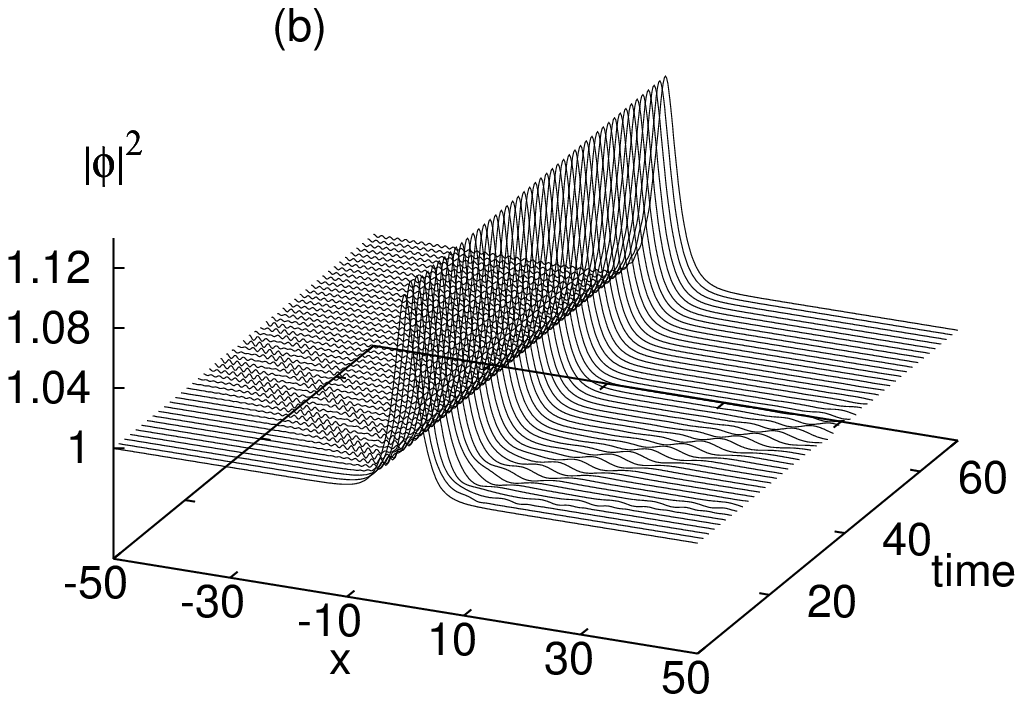}}
\caption{Time evolution of a BEC flow in (a) subcritical regime,
$v=0.373$  and (b) supercritical regime, $v=2.517$  through a
nonlinear repulsive potential barrier $V_m/\cosh(x/2)$ with
$V_m=0.5$. The initial wave packet and distribution of the BEC
local velocities $u(x)$  are taken in the form determined by
formulas (\ref{sol_u-}), (\ref{sol_u+}) and Eq. (\ref{pu1}).}
\label{repDef}
\end{figure}

Fig.~\ref{repDef} depicts time evolution of a BEC flow through a
repulsive non-linear potential $V(x)=V_m/\cosh(x/2)$ with
$V_m=0.5$ in (a) subcritical $(v=0.373<v_{-})$ and (b)
supercritical $(v=2.517>v_{+})$ regimes, respectively. Initial
form of the condensate density $\rho(x)$ is determined by
Eq.(\ref{pu1}) as $\rho(x)  = v/u(x)$, where initial distribution
of local velocities $u(x)$ is given by Eqs.~(\ref{sol_u-}),
(\ref{sol_u+}). One can see that in these regimes the flow through
the barrier is steady. Existence of small amplitude waves, 
spreading from the hump in the beginning is a result of
neglecting small terms in the course of derivation of
Eqs.(\ref{difpu1}) and (\ref{difpu2}). In
Fig.~\ref{repDef}b one can see that in supercritical regime the
 solution at the center has the hump form. The numerical
simulations  show stability of this kind of steady flows.

\begin{figure}[h]
\centerline{\includegraphics[width=8cm]{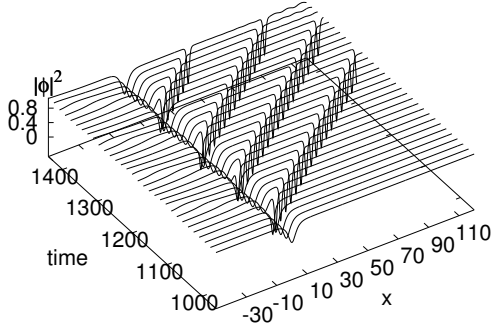}}
\caption{Time evolution of a BEC flow in the transcritical regime
when the NL barrier velocity $v=0.47$ ($v_{-}<v<v_{+}$). The NL
barrier is taken in the form of $V_m/\cosh(x/2)$ with $V_m=0.5$.
During the time period from $t=0$ to $t=1000$ (that is not
presented in the figure) the value of $V_m$ is adiabatically being
increased from $0$ to $0.5$. Further evolution is given at
$V_m=0.5$.} \label{movrepDef}
\end{figure}
In order to carry out numerical simulations of the behavior of a
BEC at transcritical velocities ($v_{-}< v < v_{+}$), we can not
use Eqs.~(\ref{sol_u-}), (\ref{sol_u+}) as an initial wave
packets, because
they have been derived for a steady flow.

In numerical simulations it is more convenient to increase
adiabatically the strength of NL potential $V_m$. In
Fig.~\ref{movrepDef} we show time evolution of BEC flow through a
NL potential barrier in the transcritical regime with $v=0.47$
$(v>v_{-})$. The NL potential is taken in the form
$V(x)=V_m/\cosh(x/2)$. $V_m$ is increasing from $0$ to $0.5$ in
the time interval $0<t<1000$ and then is kept constant. One can
see that in the transcritical regime the flow becomes unsteady and
a train of dark solitons emerges from the NL barrier at the
barrier potential strength $V_m=0.5$.


\subsection{ Short range nonlinear obstacle (delta-function potential)}

In this section we follow the approach used in the work
~\cite{Hakim}. Let us suppose the condensate to have a chemical
potential $\mu=1$. Then in the frame of the moving obstacle with
the velocity $v$ equation (\ref{eq1}) takes the form
\begin{equation}
i\psi_t + iv\psi_{x} + \frac{1}{2}\psi_{xx} - \psi - |\psi|^2\psi
= V(x)|\psi|^2\psi
 \label{eqs3}
\end{equation}
with uniform boundary conditions $|\psi(x)|^2=1$ at $x \rightarrow
\pm \infty)$.

Looking for time independent solution in the form
$\psi(x)=R(x)\exp(i\phi(x))$ we get
equations for amplitude $R(x)$ and phase $\phi(x)$
\begin{eqnarray}
\phi_x = v\left(1-\frac{1}{R^2}\right), \label{eqphi} \\
R_{xx}=v^2\left(-R+\frac{1}{R^3}\right)+R^3+V(x)R^3-R .
\label{eqR}
\end{eqnarray}
In the case of the $\delta$ function barrier potential (a sharp
jump
in the nonlinearity) $V(x)=\gamma\delta(x)$ the solution
$R(x)$ has the form
\begin{eqnarray}
R^2(x)=1  - \frac{1-v^2}{\cosh^2[\sqrt{1-v^2}(x \mp x_0)]} \
\mbox{at} \  x \lessgtr 0,  \label{R2}
\end{eqnarray}
Substituting obtained R(x) into Eq.~(\ref{eqphi}) and solving it
we obtain phase $\phi(x)$ as
\begin{eqnarray}
\phi(x) = f(x) = \ \ \ \ \ \ \  \nonumber \\
\arctan\left(\frac{2v\sqrt{1-v^2}}{\exp(\sqrt{1-v^2}(x+x_0))+2v^2-1}\right)
\ \mbox{at} \  x > 0, \  \nonumber \\
\ \mbox{and} \ \ \ \ \ \phi(x) = 2f(0)-f(-x) \ \mbox{at} \ x < 0,
\ \ \  \label{Phi}
\end{eqnarray}

where unknown parameter $x_0$ depending on the potential strength
$\gamma$ is determined from  the relation
\begin{equation}
\gamma=\frac{(1-v^2)^{3/2}\cosh(\sqrt{1-v^2}x_0)\sinh(\sqrt{1-v^2}x_0)}{\left(v^2+\sinh^2(\sqrt{1-v^2}x_0)\right)^2}
\label{x0gam}
\end{equation}
obtained from matching condition for derivatives $R_x(x)$ at $x=0$
$$R_x(+0)-R_x(-0)=\gamma R^3(0).$$
\begin{figure}[h]
\centerline{\includegraphics[width=6cm,angle=-90]{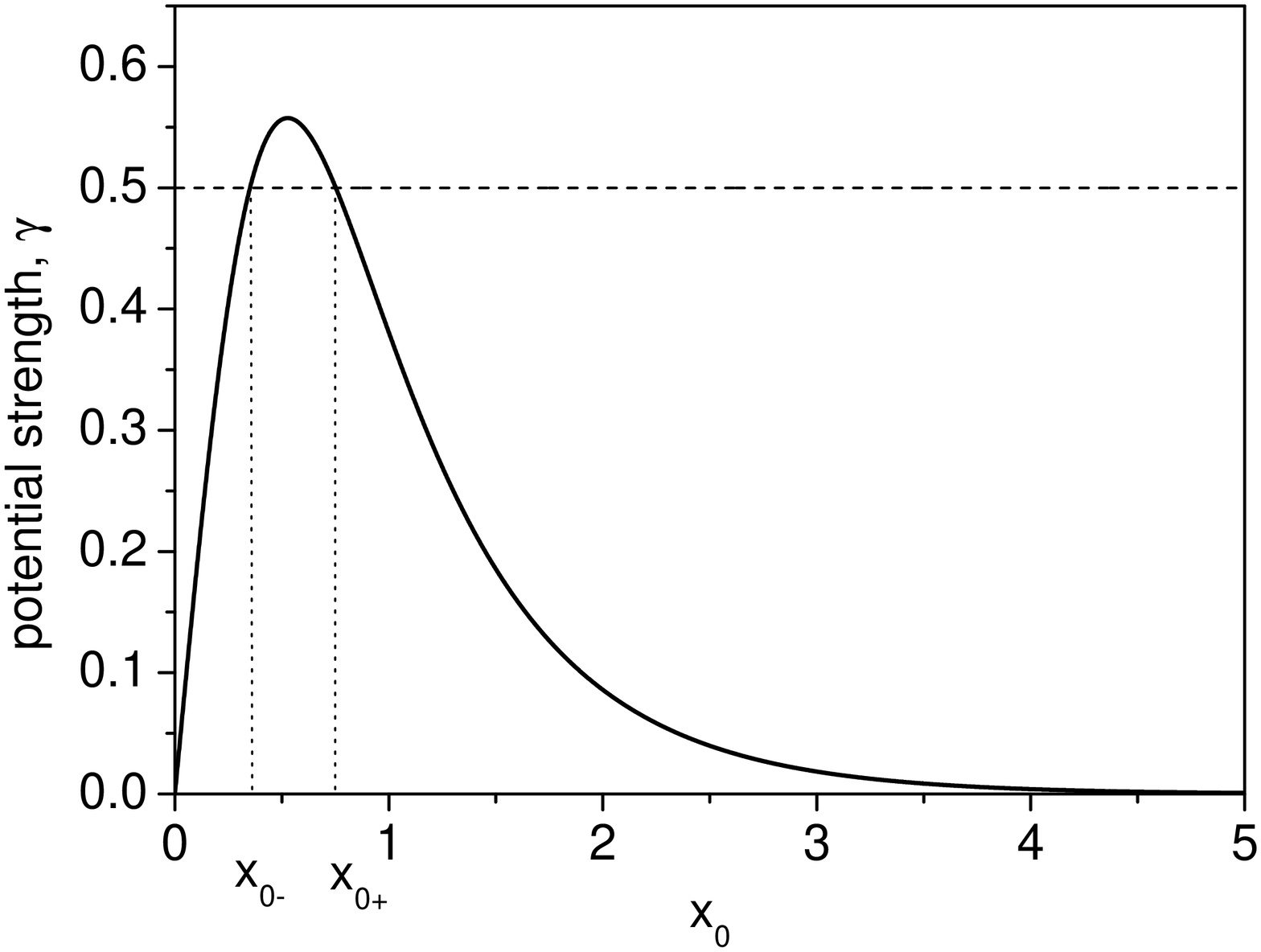}}
\caption{Dependence of the parameter $x_0$ on the nonlinear
potential strength $\gamma$ for $v=0.65$.} \label{gamma}
\end{figure}

 Fig.~\ref{gamma} depicts a typical relation between the
potential strength $\gamma$ and parameter $x_0$ at $v=0.65$. As
seen for given strength $\gamma$ there are two values of the
parameter $x_0$ (or not a single) corresponding to a pair of
steady solutions. One of the solutions ($x_0 = x_{0-}$) is
unstable and another ($x_0 = x_{0+}$) is stable
~\cite{Malomed,Hakim}.

\begin{figure}[h]
\centerline{\includegraphics[width=8cm]{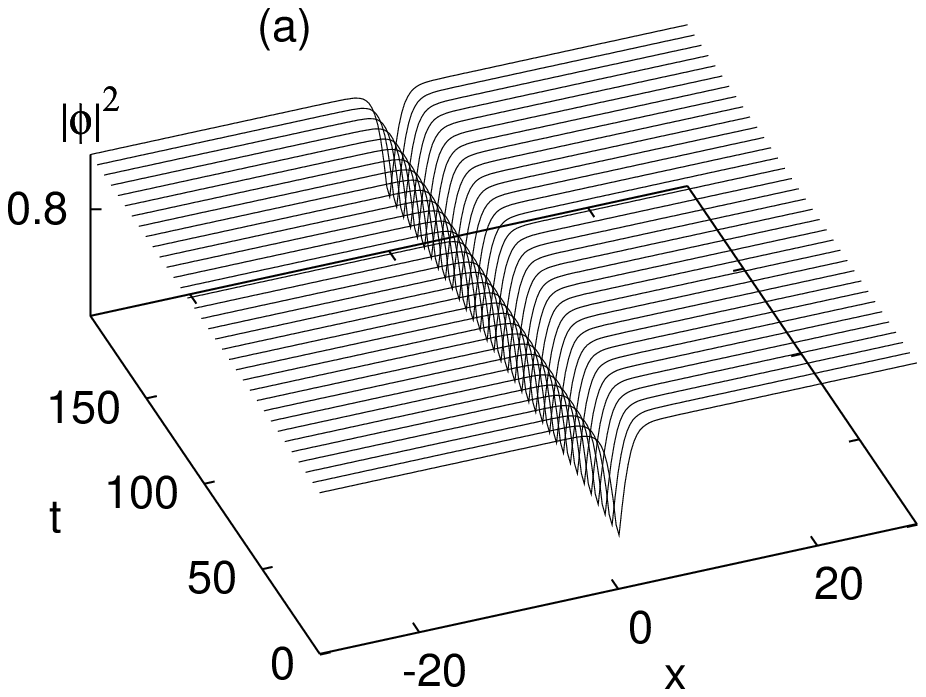}}
\centerline{\includegraphics[width=8cm]{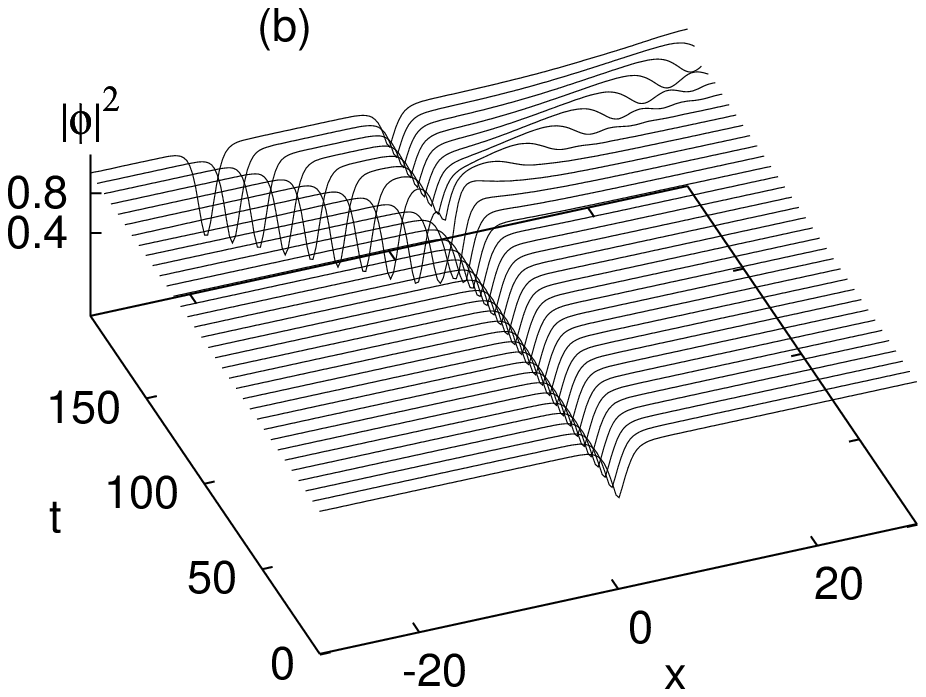}}
\caption{Time evolution of steady solutions with the parameter
$x_0=x_{0+}=0.752048$ and $x_0=x_{0-}=0.350966$ corresponding
relatively to (a) stable and (b) unstable BEC flows past a
nonlinear repulsive delta potential barrier in subcritical regime.
The other parameters  $v=0.65$ and $\gamma=0.5$. For this case
$v_{c}=0.663946$.} \label{twoSol}
\end{figure}

Time evolution of stable and unstable steady solutions
corresponding to $x_{0+}=0.752048$ and $x_{0-}=0.350966$ are shown
in Fig.~\ref{twoSol}.
 As seen the unstable solution decays into a gray soliton
moving upstream with the velocity less than $v$ and a stable
solution localized at the barrier position. The decay is
accompanied by the radiation emitted downstream in front of the
barrier.

Unlike the case of a wide barrier, in the
case of the $\delta$ function nonlinear barrier potential,
localized steady states exist only at $v<v_c<v_s$ where $v_s$ is
the sound velocity. In our case $v_s=1$. Critical velocity $v_c$
is determined by the potential strength $\gamma$
\begin{eqnarray}
\gamma=\frac{16(1-v_c^2)^2} {(6v_c^2-3+\alpha(v_c))^2}
\frac{(2v_c^2-3+\alpha(v_c))^{1/2}}
{(-2v_c^2-1+\alpha(v_c))^{1/2}},  \label{gvc}
\end{eqnarray}
where $\alpha(v_c)=\sqrt{9-4v_c^2+4v_c^4}$.

\begin{figure}[h]
\centerline{\includegraphics[width=8cm]{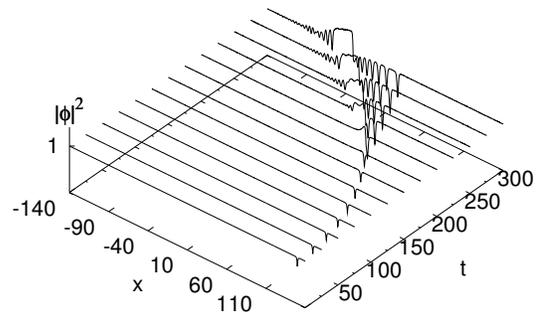}}
\caption{Time evolution of a BEC flow through the delta potential
nonlinear barrier moving with the acceleration $a=0.004$. The
barrier potential strength $\gamma=0.5$, initial velocity of the
flow $v_{0}=0$. The initial wave packet is taken in the form
Eq.~(\ref{R2}).} \label{moveNL}
\end{figure}

In order to cover a wide range of velocities we have carried out
numerical simulations of the flow of a BEC through the delta
potential nonlinear barrier moving with small acceleration
beginning from zero velocity. Fig.~\ref{moveNL} depicts the time
evolution of a BEC flow when the acceleration $a=0.004$. The
barrier potential strength $\gamma=0.5$.
The initial wave packet is taken in the form of Eq.~(\ref{R2}).

Time interval $0<t<165$ ($0<v<v_{cr}$) corresponds to a superfluid
flow. At tmes $170<t<250$ ($v_{cr}<v<v_s$) one can observe
generation of grey solitons chain.  In time interval $250<t$
($v_s<v$) corresponding to transcritical flow of a BEC at
supersonic velocity one can observe qualitatively the same wave
pattern obtained in the work~\cite{Kamchatnov} where a dispersive
shock propagates upstream with generation of soliton-like waves
propagating downstream.

\section{Conclusion}

In conclusion, we studied steady flow in a defocusing quasi 1D BEC
moving through a nonlinear {\it repulsive} barrier. Such a kind of
barriers can be formed by variation of the atomic scattering
length of BEC in {\it space}. For the case of a wide nonlinear
barrier we have found critical velocities of steady flows. Within
the interval of velocities $v_{-}<v<v_{+}$, in the transcritical
regime we observed generation of a slow moving train of dark
solitons. At velocities above supercritical the train disappears.
At the same time in this regime one can observe formation of a
hump localized at the place of {\it the barrier}.

For the case of a $\delta$ function nonlinear barrier
potential the dependence of the steady solution parameters and a
critical velocity on the
potential strength $\gamma$ was found in analytical form.  As
numerical simulations show, in subcritical regime $v<v_c$ an
unstable solution decays into a gray soliton moving upstream and a
stable solution localized at the barrier position. The decay is
accompanied by a dispersive shock  wavepropagating  downstream in front of the
barrier.

The dynamics of flows past through a linear and nonlinear barriers
are qualitatively similar except the following. In the case of a
wide {\it linear} barrier, the superfluidity is
broken at any small velocities if the barrier potential strength
greater than some threshold value (see Fig. 2 in
~\cite{Kamchatnov}). For a wide {\it nonlinear} barrier an
interval of velocities $0 < v < v_-$ {\it always} exists, where
the flow is superfluid  regardless of the barrier potential
strength.

 When using the optically induced Feshbach resonance technique
to generate a repulsive nonlinear barrier by focused laser beam,
one should in general take into account the losses, induced
by spontaneous emission of atoms. Phenomenologically it can be
described by adding a nonlinear loss term $-i\gamma|u|^2 u$ in the
GP
equation.Atom feeding can be described by linear gain term
$i\alpha u$. This case requires a separate investigation. It
should be noted that this  problem relates to one considered in
the recent work~\cite{KK12}, where the flow of polariton
condensate~\cite{amo} past a linear barrier was studied taking
into account linear amplification and nonlinear damping.

\begin{acknowledgments}
Authors are grateful to E.N. Tsoy for fruitful discussions.
F.Kh.A. acknowledges a  Marie Curie Grant No.PIIF-GA-2009-236099(NOMATOS).
\end{acknowledgments}


\newpage

\begin{thebibliography}{99}

\bibitem{Kamchatnov}
A.M. Leszczyszyn, G.A. El, Yu.G. Gladush, and A.M. Kamchatnov,
Phys.Rev. A {\bf 79}, 063608 (2009).

\bibitem{AP04}
G.E. Astrakharchik and L.P. Pitaevskii, Phys.Rev. A {\bf 70},
013608 (2004).

\bibitem{EA07}
P.Engels and C. Atherton, Phys.Rev.Lett. {\bf 99},160405 (2007).

\bibitem{Malomed}
B. Malomed and M. Azbel, Phys.Rev. B {\bf 47}, 10402 (1993).

\bibitem{Hakim}
V. Hakim, Phys.Rev. E {\bf 55}, 2835 (1997).

\bibitem{Pitaevskii}
L. P. Pitaevskii and S. Stringari, Phys. Lett. A  {\bf 235}, 398 (1997).

\bibitem{Feynman}
R.P. Feynman, in {\it Progress in Low Temperature Physics}, edited by C.J. Coster (North-Holland, Amsterdam, 1955), Vol.I, p.17.

\bibitem{Law}C. K. Law, C. M. Chan, P. T. Leung, and M.-C. Chu, Phys. Rev. Lett. {\bf 85}, 1598 (2000).

\bibitem{Hakim2}M. Haddad and V. Hakim, Phys. Rev. Lett {\bf 87}, 218901 (2001).

\bibitem{Pavloff}
N. Pavloff, Phys.Rev. A {\bf 66}, 013610 (2002).

\bibitem{Inouye}
S. Inouye et al., Nature {\bf 392}, 151 (1998).

\bibitem{Schlyapnikov}
P.O. Fedichev, Yu. Kagan, G.V. Schlyapnikov, and J.T.M. Walraven, Phys.Rev.Lett. {\bf 77}, 2913 (1996).

\bibitem{NOLExp2}
R. Yamazaki, S. Taie, S. Sugawa, and Y. Takahashi, Phys. Rev. Lett. {\bf 105},
050405 (2010).

\bibitem{BG10}
N. Berloff and V.M. Perez-Garcia, arXiv:1006.4426.

\bibitem{KK12}
A.M. Kamchatnov and Y.V. Kartashov, EPL {\bf 97}, 1006 (2012).

\bibitem{amo}
A. Amo et al., Science, {\bf 332}, 1167 (2011).
\end{thebibliography}
\end{document}